\documentstyle[epsfig]{aipproc}
\newcommand{\apj}[1]{{\it Astrophys.\ J. }{\bf #1}}
\newcommand{\mn}[1]{{\it Mon.\ Not.\ Roy.\ Astron.\ Soc.\ }{\bf #1}}
\newcommand{\prd}[1]{{\it Phys.\ Rev.\ D }{\bf #1}}
\newcommand{\prl}[1]{{\it Phys.\ Rev.\ Lett.\ }{\bf #1}}

\begin{document}

\title{Gravitational waves from the $r$-modes of rapidly rotating
  neutron stars}
\author{Benjamin J. Owen}
\address{Max Planck Institut f\"ur Gravitationsphysik (Albert Einstein
  Institut) \\ Am M\"uhlenberg 1, 14476 Golm bei Potsdam, Germany}
\maketitle

\begin{abstract}
  Since the last Amaldi meeting in 1997 we have learned that the
  $r$-modes of rapidly rotating neutron stars are unstable to
  gravitational radiation reaction in astrophysically realistic
  conditions.  Newborn neutron stars rotating more rapidly than about
  100~Hz may spin down to that frequency during up to one year after
  the supernova that gives them birth, emitting gravitational waves
  which might be detectable by the enhanced LIGO interferometers at a
  distance which includes several supernovae per year.  A cosmological
  background of these events may be detectable by advanced LIGO.  The
  spins (about 300~Hz) of neutron stars in low-mass x-ray binaries may
  also be due to the $r$-mode instability (under different
  conditions), and some of these systems in our galaxy may also
  produce detectable gravitational waves---see the review by G.
  Ushomirsky in this volume.  Much work is in progress on developing
  our understanding of $r$-mode astrophysics to refine the early,
  optimistic estimates of the detectability of the gravitational
  waves.
\end{abstract}

\section{The $r$-Mode Instability}

The $r$-mode instability has been the subject of about thirty papers
over the past two years.  I will not be able to do them all justice
here.\footnote{For a recent review with more emphasis on completeness,
  see Friedman and Lockitch \cite{fl}.}  Instead I will summarize the
most important (as I see them) results with a direct impact on
gravitational-wave detection, beginning with the basic model worked
out in 1998 and ending with the latest (end of 1999) developments in
this rapidly changing field.

The reason for the excitement is a version of the CFS
instability---named for Chandrasekhar, who discovered it in a special
case \cite{c70}, and for Friedman and Schutz, who investigated it in
detail and found that it is generic to rotating perfect fluids
\cite{fs78}.  The CFS instability allows some oscillation modes of a
fluid body to be driven rather than damped by radiation reaction,
essentially due to a disagreement between two frames of reference.

The mechanism can be explained heuristically as follows.  In a
non-rotating star, gravitational waves radiate positive angular
momentum from a forward-moving mode and negative angular momentum from
a backward-moving mode, damping both as expected.  However, when the
star rotates the radiation still lives in a non-rotating frame.  If a
mode moves backward in the rotating frame but forward in the
non-rotating frame, gravitational radiation still removes positive
angular momentum---but since the fluid sees the mode as having
negative angular momentum, radiation drives the mode rather than damps
it.  Another example of such an effect due to a disagreement between
frames of reference is the well-known Kelvin-Helmholtz instability,
which leads to rough airplane rides over the jet stream and pounding
surf on the California coast.

Mathematically, the criterion for the CFS instability is
\begin{equation}
\label{crite}
\omega (\omega+m\Omega) < 0,
\end{equation}
with the mode angular frequency $\omega$ (in an inertial frame) in
general a function of the azimuthal quantum number $m$ and rotation
angular frequency $\Omega$.  For any set of modes of a perfect fluid,
there will be some modes unstable above some minimum $m$ and $\Omega$.
However, fluid viscosity generally grows with $m$ and there is a
maximum value of $\Omega$ (known as the Kepler frequency $\Omega_K$)
above which a rotating star flies apart.  Therefore the instability is
only astrophysically relevant if there is some range of frequencies
and temperatures (viscosity generally depends strongly on temperature)
in which it survives.

The $r$-modes are a set of fluid oscillations with dynamics dominated
by rotation.  They are in some respects similar to the Rossby waves
found in the Earth's oceans and have been studied by astrophysicists
since the 1970s \cite{pp78}.  The restoring force is the Coriolis
inertial ``force'' which is perpendicular to the velocity.  As a
consequence, the fluid motion resembles (oscillating) circulation
patterns. The (Eulerian) velocity perturbation is
\begin{equation}
\label{dv}
\delta\vec{v} = \alpha\Omega R(r/R)^m \vec{r} \times \vec{\nabla}
Y_{mm}(\theta,\phi) +O(\Omega^3),
\end{equation}
where $\alpha$ is a dimensionless amplitude (roughly $\delta v/v$) and
$R$ is the radius of the star.  Since $\delta\vec{v}$ is an axial
vector, mass-current perturbations are large compared to the density
perturbations.  The Coriolis restoring force guarantees that the
$r$-mode frequencies are comparable to the rotation frequency,
\begin{equation}
\omega+m\Omega = {2\over m+1} \Omega +O(\Omega^3).
\end{equation}
It was not until the time of the last Amaldi Conference in mid-1997
that Andersson~\cite{a98} noticed that the $r$-mode frequencies
satisfy the mode instability criterion~(\ref{crite}) for all $m$ and
$\Omega$, and that Friedman and Morsink~\cite{fm98} showed the
instability is not an artifact of the assumption of discrete modes but
exists for generic initial data.  In other words, {\it all} rotating
perfect fluids are subject to the instability.

%\begin{figure}
%\centerline{\epsfig{file=rmode2.ps,height=3.5in,width=3.5in}}
%\caption{The quadrupolar $r$-mode. Velocity pattern and fluid motion.}
%\label{f:r-mode}
%\end{figure}

\section{Driving vs.\ Damping}

The universe is inhabited not by balls of perfect fluid, but by stars
subject to internal viscous processes which tend to damp out
oscillation modes.  To evaluate the stability of modes in realistic
neutron stars, we must compare driving and damping timescales.

In the small-amplitude limit, a mode is a driven, damped harmonic
oscillator with an exponential damping timescale \cite{il91}
\begin{equation}
{1\over\tau} = -{1\over2E} {dE\over dt} = -{1\over2E} \left[
\left(dE\over dt\right)_G +\sum_V \left(dE\over dt\right)_V \right]
= {1\over\tau_G} +\sum_V {1\over\tau_V} .
\end{equation}
Here $E$ is the energy of the mode in the rotating frame and $dE/dt$
is the sum of contributions from gravitational radiation (subscript
$G$) and all viscous processes (subscript $V$).  The mode is stable
if the damping timescale $\tau$ is positive, unstable if $\tau$ is
negative.  The gravitational radiation timescale $\tau_G$ depends on
the rotation frequency $\Omega$, and the viscous timescales generally
depend also on the temperature $T$.  Therefore we define a critical
frequency $\Omega_c$ such that
\begin{equation}
{1\over \tau(\Omega_c,T)} = 0
\end{equation}
and decide if a given mode is astrophysically interesting by examining
the curve $\Omega_c(T)$.

Neutron stars are complicated objects, but a simple model suffices to
estimate the most important driving and damping timescales in the very
young ones.  When hotter than $10^9$K (younger than about a year),
most of the star is a ball of ordinary, barotropic (equation of state
independent of temperature) fluid.  Given a putative equation of
state, the gravitational radiation timescale $\tau_G$ can be
calculated by standard multipole integrals~\cite{t80}, although the
$r$-modes are nonstandard in that the leading-order (in $\Omega$)
contribution is not from the mass multipoles but from the mass-current
multipoles~\cite{lom98}.  Viscous damping is due both to shearing of
the fluid and to compression and rarefaction of individual fluid
elements (bulk viscosity).  The shear viscosity is stronger (timescale
is shorter) at lower temperatures (like everyday experience with motor
oil) and can be calculated from neutron-neutron scattering
cross-sections~\cite{cl87}.  The bulk viscosity is a weak nuclear
interaction effect and thus is much stronger at higher temperatures.
Compression and rarefaction of the fluid by the mode disturbs the
density-dependent equilibrium $p+e \leftrightarrow n$, generating
neutrinos which efficiently carry energy away~\cite{s89}.  As the star
cools the viscous mechanisms change (see Sec.~\ref{open}), but this
model is good enough for a first look.

\begin{figure}[t]
\centerline{\epsfig{file=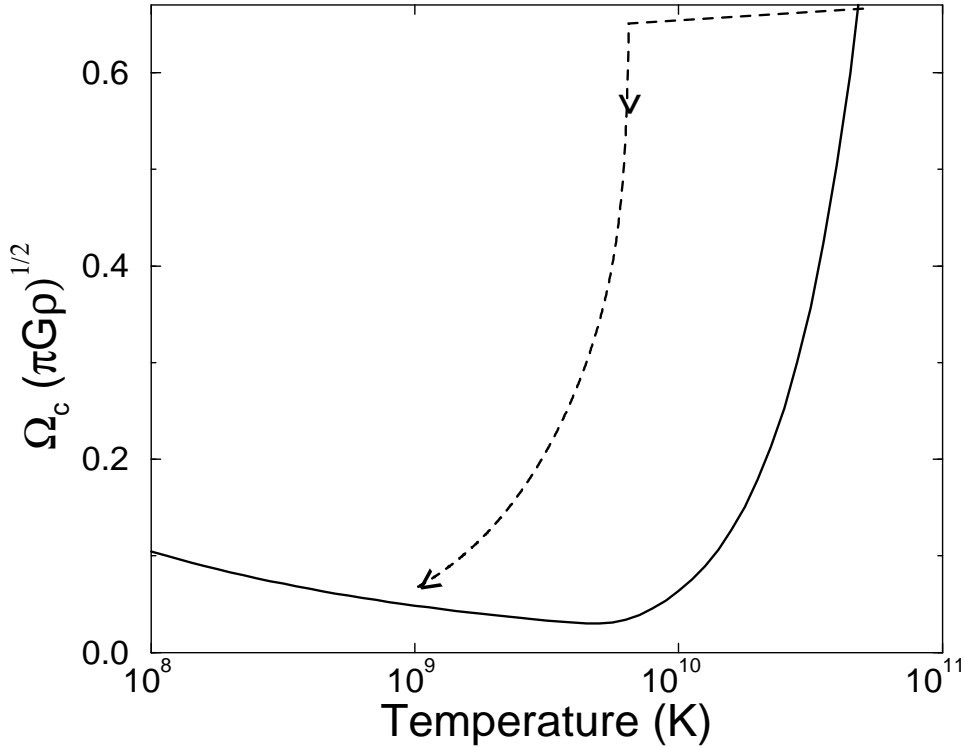,height=4.in,width=5.in}}
\vspace{10pt}
\caption{The solid curve is critical angular velocity $\Omega_c$ as a
  function of temperature $T$, assuming ordinary fluid viscosity given
  by Lindblom, Mendell, and Owen~\protect\cite{lmo99}.  Above this
  curve, the fastest-growing ($m=2$) $r$-mode is unstable.  The dashed
  curve is the evolutionary track of a neutron star in the first year
  of its life if the $r$-modes begin at amplitude $\alpha=10^{-5}$ and
  saturate at $\alpha=1$, although it is not very sensitive to these
  values.}
\label{f:crit}
\end{figure}

The net damping timescale of the most unstable ($m=2$) $r$-mode can be
written in terms of fiducial timescales (written with tildes)
\begin{equation}
{1\over \tau(\Omega,T)} = {1\over \tilde\tau_G} {\Omega^6\over
(\pi G\bar\rho)^3} +{1\over \tilde\tau_S} \left(10^9\mbox{K}\over
T\right)^2 +{1\over \tilde\tau_B} \left(T\over 10^9\mbox{K}\right)^6
{\Omega^2\over \pi G\bar\rho},
\end{equation}
where $\bar\rho$ is the mean density of the equilibrium star.  The
numerical values of the fiducial timescales (for a simplistic equation
of state) have been evaluated as~\cite{lom98,aks99,lmo99}
\begin{equation}
\tilde\tau_G = -3.3\mbox{ s}, \quad \tilde\tau_S = 2.5\times10^8
\mbox{s}, \quad \tilde\tau_B = 2.0\times10^{11}\mbox{s} .
\end{equation}
The numbers change by factors of two or so for different neutron-star
models, but the curve plotted in Fig.~\ref{f:crit} and its conclusion
are very robust.  The $r$-modes are unstable in realistic neutron
stars over an interesting range of frequencies and temperatures.
Neutron stars born rotating at or near the Kepler frequency will spin
down and emit gravitational radiation in the process; the question now
is how much.

\section{Spindown-Cooling Model}
\label{model}

To detect gravitational waves from the $r$-modes, we need to know how
the modes grow beyond the limits of perturbation theory and spin down
the neutron star as it cools during the first year of its life.  Even
in the simple approximation of an ordinary fluid ball, this involves
nonlinear hydrodynamics and radiation reaction, which are both tricky
subjects and could take years to explore properly.  In the meantime we
make do with a simple model~\cite{o-98} developed to make the first,
rough estimates of detectability.

In this model, we consider three coupled systems---a uniformly
rotating fluid background (with angular velocity $\Omega$), the most
unstable $r$-mode (with dimensionless amplitude $\alpha$), and the
rest of the universe.  The two systems in the star (mode and
background) couple to each other by viscosity and nonlinear fluid
effects.  The mode couples to the universe by gravitational radiation;
the background does not.  The mode's energy evolves by gravitational
radiation and viscous damping; the behavior of the background is
determined by conservation of energy and angular momentum.  Although
some of the mode's energy goes into heating the star, the standard
neutrino cooling law ($T/10^9$K) $\sim$ (1~yr$/t$)$^{1/6}$ is all but
unaffected.

If a star is born spinning at $\Omega_K$ at temperature $10^{11}$K
(and recent observations~\cite{fast} suggest that some stars are), the
evolution falls into three distinct phases.  The {\it growth phase}
begins when the $r$-modes go unstable of order one second after the
supernova.  During this phase a small initial perturbation $\alpha$
grows exponentially on a timescale of order one minute while $\Omega$
remains almost constant (the mode is too small to emit much angular
momentum).  In this regime linearized hydrodynamics (which is all we
know at the moment) is a good approximation.  Within at most a few
minutes after the supernova, $\alpha$ becomes so large that nonlinear
hydrodynamic effects can no longer be neglected.  Previous studies of
other modes~\cite{dl77} indicate that the main effect might be a
saturation of the mode amplitude at some constant value, which we can
treat as a phenomenological parameter.  In this {\it saturation
  phase}, the star spins down very rapidly ($d\Omega/dt \sim
\Omega^7$) and emits gravitational radiation of strain amplitude
\begin{equation}
h(t) = 4\times10^{-24} \left(\Omega \over \pi G\bar{\rho} \right)^3
\left(20\mbox{ Mpc} \over D\right) \alpha_{\rm max},
\end{equation}
at a detector at distance $D$ (normalized here to the distance at
which we expect several events per year).  As the star spins down, the
gravitational radiation gets much weaker (recall $1/\tau_G \sim
\Omega^6$).  Also, viscous damping becomes stronger, especially since
other mechanisms come into play---for instance, when the neutrons
become superfluid after cooling to about $10^9$K.  Thus, within a year
the star has moved along a track such as that in Fig.~\ref{f:crit} and
entered the {\it decay phase}, where the $r$-modes are stabilized by
viscosity and $\alpha$ slowly dies away without changing $\Omega$
much.  The final spin frequency $\Omega_{\rm end}$ is in practice
another phenomenological parameter, since it depends on the more
complicated viscous processes of cooler neutron stars as well as on
$\alpha_{\rm max}$.

\section{Detectability of Gravitational Waves}

Even at its strongest, an $r$-mode signal is below the strain noise
of a gravitational-wave detector.  But electromagnetic astronomers
have been pulling faint pulsar signals out of noisy data for decades,
and their data analysis techniques can be adapted for the $r$-modes.

\begin{figure}[t]
\centerline{\epsfig{file=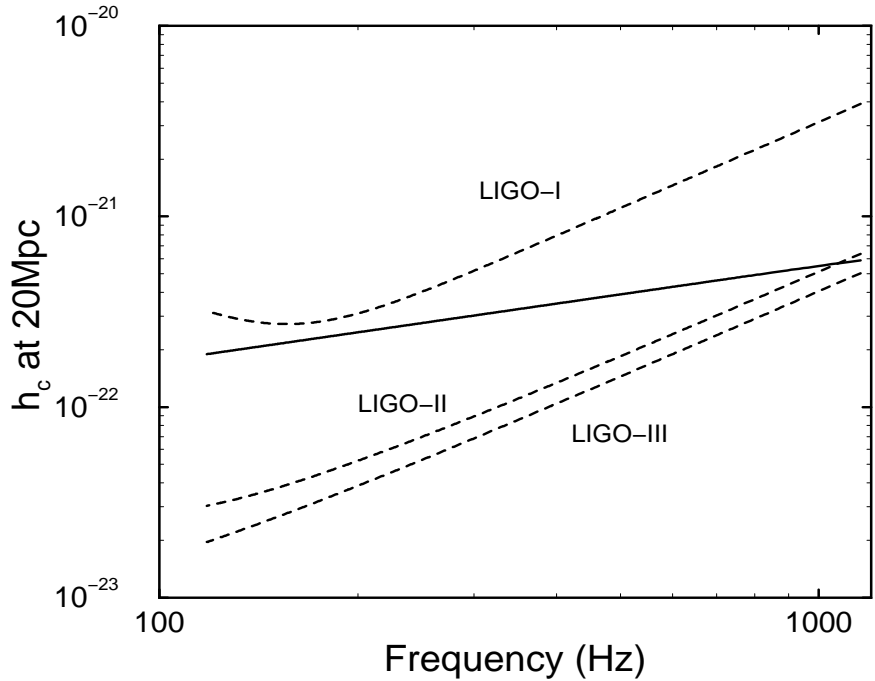,height=4.in,width=5.in}}
\vspace{10pt}
\caption{Characteristic signal amplitude $h_c$ and noise $h_{\rm
rms}$.  The signal is from a neutron star with mass $1.4M_\odot$ and
radius 12.5~km at a distance of 20~Mpc (several events per year).  The
noise is for the three versions of LIGO used in
Ref.~\protect\cite{o-98}.  Integrated (optimal) signal-to-noise ratio
is 8; realistic signal-to-noise is somewhat less.}
\label{f:hc}
\end{figure}

Surprisingly, even the crude model of the source given in
Sec.~\ref{model} is good enough to estimate the detectability of the
gravitational waves.  The quantity of interest is not the raw strain
$h(t)$ but rather a {\it characteristic strain}
\begin{equation}
h_c(f) = h[t(f)] \sqrt{f^2/|df/dt|},
\end{equation}
where $df/dt$ is the time derivative of the gravitational wave
frequency.  The optimal (filtered) signal-to-noise ratio is
\begin{equation}
(S/N)^2 = 2\int (d\ln{f}) (h_c/h_{\rm rms})^2,
\end{equation}
where the rms strain noise is related to the detector's power
spectral noise density by
\begin{equation}
h_{\rm rms} = \sqrt{f\,S_h(f)}.
\end{equation}
Thus $(S/N)^2$ can be estimated by looking at a plot such as
Fig.~\ref{f:hc}.  For the $r$-modes, we find that~\cite{o-98}
\begin{equation}
h_c = 6\times10^{-22} \left(f \over \mbox{1 kHz} \right)^{1/2}
\left(20\mbox{ Mpc} \over D\right)
\end{equation}
with $(S/N)=8$ for the projected LIGO-II (enhanced) noise curve as of
1998.  This result is independent of much of the detailed physics of
the source, including $\alpha_{\rm max}$.\footnote{To my knowledge the
  argument was first made by R. D. Blandford in 1984 (but never
  published) that such a robust result holds for any system evolving
  mainly via gravitational radiation---like the $r$-modes in the
  saturation phase.}  However, it does depend on the detailed
astrophysics through the final low-frequency cutoff, which does not
change much even if the viscous damping changes by orders of
magnitude.

The optimal signal-to-noise ratio is only an upper limit---it assumes
matched filtering, which requires precise tracking of the signal
phase.  While our knowledge of astrophysics will never be good enough
to track an $r$-mode signal to within one cycle out of $10^9$, there
are alternatives.  The lower limit has been set by Brady and
Creighton~\cite{bc} using the simplest possible search algorithm,
patterned on the techniques used to find pulsar signals in
electromagnetic data.  Assuming the supernova has been observed
optically and a sky position is available, the Doppler shifts due to
the Earth's motion can be removed to obtain a signal which is
sinusoidal but for the (slow) intrinsic frequency evolution of the
source.  Even without any modeling of this evolution, i.e.\ by
expanding
\begin{equation}
f(t) = f_0 \left( 1+ \sum_{k} f_k t^k \right)
\end{equation}
for short Fourier transforms (integrating for year is computationally
too expensive) and combining the transforms in some way for different
trial values of the spindown parameters $f_k$, it is possible to
obtain one fifth of the optimal signal-to-noise.  This is in spite of
the fact that the search is computationally limited by the requirement
that data analysis keep pace with data acquisition and by the fact
that the $r$-modes evolve so quickly that many terms $f_k$ are needed.
With constraints---even rough ones---from a physical model, the $f_k$
are no longer all independent and the efficiency of data analysis
could be increased.  Since event rate goes roughly as $(S/N)^3$, it is
important to beat the lower limit.

A stochastic background from the superposition of many faint $r$-mode
signals out to cosmological distances will also exist.  However, it is
much fainter than a single signal and thus detectable only by
(advanced) LIGO-III~\cite{o-98,fms99}.

\section{Open Questions}
\label{open}

We are now (at the end of 1999) in the midst of a renewed flurry of
activity on $r$-mode astrophysics.  Several effects neglected in the
first simple scenario are being worked out.  Some of them could damp
the $r$-modes much more effectively than previously thought, pushing
the detectability of the gravitational waves from LIGO-II to LIGO-III.
However, this is far from certain and the astrophysicists are having
an exciting time working it out.  Here is a list of the effects that
(I think) have the most direct impact on detection prospects.

{\it Superfluid viscosity.}  One of the most eagerly awaited papers
has been the calculation of the damping effects of ``mutual
friction'', a process which paradoxically increases the viscous
damping when a neutron star cools to a superfluid.  At temperatures
below about $10^9$K this viscous mechanism was expected to dominate,
and the big question was whether the damping was sufficient to
stabilize the $r$-modes in stars older than about a year---especially
the low-mass x-ray binaries (see the review by G. Ushomirsky in this
volume).  The answer~\cite{lm} is a definite maybe.  The damping
timescale varies by several orders of magnitude, depending on a
parameter of superfluid physics (the neutron-proton entrainment
coefficient) which is yet poorly known.  More work is in progress,
but recently mutual friction has been upstaged by other issues.

{\it Relativistic effects.}  Most $r$-mode calculations to date have
assumed Newtonian gravity.  Relativity was thought to simply multiply
various numbers by redshift factors of order unity, but there are two
important qualitative differences with Newtonian gravity.  First there
is the claim by Kojima \cite{k98} that the $r$-mode frequency becomes
smeared over a finite bandwidth.  This claim is contradicted, however,
by Lockitch \cite{l99}.  With Andersson and Friedman, he \cite{alf}
finds that relativistic $r$-modes do however have an increased
coupling to bulk viscosity similar to that of the ``generalized
$r$-modes'' \cite{li99,lf99} of Newtonian stars, which are still
unstable but less so.

{\it Nonlinear fluid dynamics.}  At least two groups \cite{r-,fsk} are
working on codes to numerically solve the fully nonlinear fluid
equations for the $r$-modes and determine the saturation amplitude.
However, the problem is complicated and the investment of coding and
formalism is large, so expect results in a year or two at best.  Order
of magnitude arguments~\cite{small} have been made to claim that the
$r$-modes saturate due to mode-mode coupling at a very small amplitude
$\alpha \sim 10^{-5}$, which would render the signal undetectable.
However, these arguments neglect the unique symmetries of the
$r$-modes; and based on work on the $g$-modes of white dwarfs
\cite{gw99} it seems that the $r$-modes could indeed grow much larger.
Semi-analytical analyses~\cite{mmc} of mode-mode coupling may give
some indications about mode saturation while we wait for numerical
results.

{\it Magnetic fields.}  If the growth of the $r$-modes leads to
substantial differential rotation, it could wind up magnetic field
lines frozen into the fluid of a young neutron star, amplifying any
seed field and saturating the $r$-modes at a small amplitude.  Two
recent papers \cite{rls,lu} claim that the $r$-modes produce
differential rotation, but this is a nonlinear effect which the
authors have tried to treat with linear perturbation theory.  Strictly
speaking the gravitational radiation is also nonlinear (quadratic in
$\alpha$), but the canonical energy and angular momentum are global
quantities whose perturbation can be derived self-consistently from a
Lagrangian principle.  It is not clear how to do this for local,
dynamical quantities such as vorticity.

{\it Crust formation.}  Perhaps the most important new result is that
the formation of a solid crust (below about $10^{10}$K) can act to
strongly stabilize the mode.  Bildsten and Ushomirsky~\cite{bu} find
that shear viscosity in the fluid boundary layer just below the crust
decreases the damping timescale by $10^5$--$10^7$.  They conclude that
the $r$-modes are completely suppressed in low-mass x-ray binaries and
that the signal-to-noise ratio is reduced by three for newborn neutron
stars.  But it is not clear to me that this result is correct for
newborn neutron stars.  If an $r$-mode is already excited when the
crust starts to form (of order a minute after the supernova), the
intense and localized shear heating in the boundary layer can re-melt
the crust if the pre-existing $r$-mode is strong enough.  In this
case, the outer layers stay in a self-regulating equilibrium at the
melting temperature and the old model of the evolution is largely
unaffected.  I estimate that, in this case, ``strong enough'' means an
$r$-mode amplitude of $\alpha=10^{-3}$.  This points out some
interesting questions for future research: First, what is the initial
value of $\alpha$ when the $r$-modes first go unstable?  The first
model~\cite{o-98} used gratuitously small values to make a point, but
no one knows yet what are reasonable values.  Also, what exactly is
the melting temperature of a new crust?  If it is $8\times10^9$K
rather than $10^{10}$K then the $r$-mode could have plenty of time to
grow, and in astrophysics a 20\% error is considered high precision.

Although I have skipped over many astrophysics issues, I realize even
this short list may be bewildering to the experimenters and data
analysts who are the main audience at the Amaldi Conference.  If I had
to distill my presentation into one sentence, I would say: Let the
theorists argue for another two years; the $r$-modes are not as good a
bet as binaries, but they may not be far behind.

\section{Acknowledgments}

I am grateful to many colleagues for discussions of published and
especially unpublished work which enabled me to give a good review:
H. Asada, \'E. Flanagan, J.-A. Font, J. Friedman, J. Ipser, F. Lamb,
Y. Levin, K. Lockitch, G. Mendell, S. Morsink, E. S. Phinney,
L. Rezzolla, N. Stergioulas, K. Thorne, G. Ushomirsky, and especially
L. Lindblom.

\end{document}